\documentclass[twocolumn,prl]{revtex4}
\usepackage{graphicx}    
\usepackage{dcolumn}     
\usepackage{bm}          
\newcommand{\Vec}[1]{\mbox{\boldmath$#1$}}
\begin{document}
\title{A unified origin for the 3D magnetism and 
superconductivity in Na$_x$CoO$_2$}
\author{
Kazuhiko Kuroki$^1$, Shuhei Ohkubo$^{1*}$, 
Takumi Nojima$^{1**}$,Ryotaro Arita$^2$, 
Seiichiro Onari$^3$, and Yukio Tanaka$^3$
}
\affiliation{
$^1$ Department of Applied Physics and Chemistry, 
The University of Electro -Communications, Chofu, Tokyo 182-8585, Japan\\  
$^2$ RIKEN, 2-1 Hirosawa, Wako, Saitama 351-0198, Japan\\
$^3$ Department of Applied Physics, Nagoya University, Nagoya, 464-8603, 
Japan}
\date{\today}
\begin{abstract}
We analyze the origin of the three dimensional (3D) magnetism observed in 
nonhydrated Na-rich Na$_x$CoO$_2$ within an itinerant spin 
picture using a 3D Hubbard model. The origin is identified 
as the 3D nesting between the inner and outer portions of the Fermi surface,
which arise due to the local minimum structure of the $a_{1g}$ band 
at the $\Gamma$-A line. The calculated spin wave dispersion strikingly 
resembles the neutron scattering result. 
We argue that this 3D magnetism and the 
spin fluctuations responsible for superconductivity in the hydrated systems 
share essentially the same origin.
\end{abstract}
\pacs{PACS numbers: }
\maketitle

The discovery of superconductivity (SC) in 
Na$_x$CoO$_2\cdot y$ H$_2$O has attracted much attention.\cite{Takada}
Although the experiments are still somewhat controversial, 
we briefly summarize the present understanding.
(i) Up to now, all the angle 
resolved photoemission (ARPES) measurements
\cite{Hasan,Yang2,Takeuchi,Shimojima} 
show the absence of the 
$e_g'$ hole pockets predicted in the first principles calculation.\cite{Singh}
Thus the remaining $a_{1g}$ seems to be the relevant band.
(ii) In the bilayer
hydrated (BLH) SC samples, there is an enhancement in $(T_1T)^{-1}$ 
at low temperatures at the Co site
\cite{Ishida3,Fujimoto,Imai,Ihara,Michioka} and at the O site,
\cite{Imai,Ihara2} 
($T_1^{-1}$ is the spin-lattice relaxation rate)   
indicating the presence of spin fluctuations (SF) located away from the 
Brillouin zone (BZ) edge,\cite{Imai,Ihara2}
while such an enhancement is not seen in non-SC monolayer hydrated and 
nonhydrated Na-poor samples. 
On the other hand, the Knight shift stays nearly constant at 
low temperatures in the SC samples,\cite{Imai,Ihara2,Alloul}  
which means that the SF is not purely ferromagnetic.
The bottom line 
is that SF that is  
neither purely ferromagnetic nor antiferromagnetic 
is strongly related to SC. 
(iii) Unconventional SC gap is suggested from the absence of the 
coherence peak in $T_1^{-1}$ as well as the 
power law decay below $T_c$.\cite{Ishida3,Fujimoto,Kobayashi,Zheng} 
Several experiments suggest 
spin singlet pairing,\cite{Kobayashi,Zheng2} 
and 
moreover, the effect of 
the impurities on $T_c$ is found to be small, \cite{Yokoi,Oeschler} 
suggesting a $s$-wave like gap.

In a previous paper, as a solution for understanding 
these experimental results, 
we have proposed a mechanism for an unconventional 
$s$-wave pairing,  
in which the local minimum structure (LMS) of the $a_{1g}$ band at the 
$\Gamma$ point plays an important role.\cite{Kuroki} 
For appropriate band fillings, 
LMS results in 
disconnected inner and outer Fermi surfaces (FS),
whose partial nesting 
induces SF at wave vectors that bridge the
 two FS (see Fig.\ref{fig6} left panel).
This incommensurate SF can give rise to 
an ``extended $s$-wave'' pairing 
in which the gap changes sign between the two FS, but 
not within each FS. 
A recent multiorbital analysis also shows that 
such a pairing can occur when the inner and outer 
$a_{1g}$ FS are present.\cite{MO}

The purpose of this paper is to investigate the relation between this 
SC and the 3D  spin-density-wave(SDW)-like, metallic 
magnetism observed in the nonhydrated Na-rich systems, 
\cite{Motohashi,Sugiyama,Foo,Keimer1}  
whose magnetic structure has been revealed by 
neutron scattering experiments to have 
in-plane ferromagnetic and out-of-plane antiferromagnetic character.
\cite{Keimer2,Helme}
Neutron scattering experiments have further obtained the spin wave 
dispersion of this 3D magnetism. Analysis of this dispersion 
based on the localized spin 
picture have found that the out-of-plane antiferromagnetic coupling 
is of the order of the in-plane ferromagnetic coupling,
\cite{Keimer2,Helme,Johannes} 
despite the strong 2D nature of the material.
Moreover, Curie-Weiss temperature is expected to be positive from the 
evaluated coupling constants,\cite{Keimer2} 
while it is actually negative.
\cite{Motohashi,Sugiyama,Foo,Keimer1,Galivano,Wang,Alloul}
A theory based on a spin-orbital polaron picture
has been proposed for this puzzle.\cite{Khaliullin}

In the present study, since the 
system remains metallic in the magnetically ordered state, 
we take an itinerant spin viewpoint. From the consistency between the 
experiments and the calculated results, we propose that the origin of 
the 3D magnetism is essentially the same with 
the SF responsible for SC : the nesting between 
inner and outer portions of the Fermi surface that arises due to the 
local minimum of the $a_{1g}$ band.\cite{Korshunov}

\begin{figure}[t]
\begin{center}
\includegraphics[width=8cm,clip]{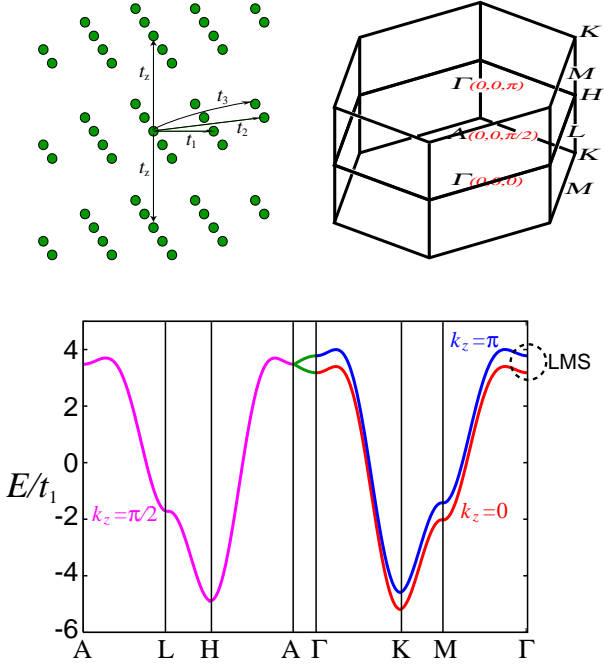}
\caption{(color online) 
Upper left: the 3D model in the present study. Upper right: the 
Brillouin zone. The wave numbers in the parenthesis are those in the unfolded 
BZ scheme. Lower panel: the band structure of the present model 
with $U=0$. 
$k_z=0,\pi/2,\pi$ 
are $k_z$ in the unfolded BZ scheme.
\label{fig1}}
\end{center}
\end{figure}

There exist two 
CoO$_2$ layers within a unit cell due to the alternation of the 
oxygen arrangement, but neither the effective hopping integrals 
nor the on-site energy alter along the $c$ axis, 
so the  BZ folding in the $c$ direction 
occurs without inducing a gap in the band. Thus, in a 3D effective theory 
in which O and Na site degrees of freedom are integrated out, 
the unit cell contains only one layer, which results in an 
unfolded BZ shown in the upper right of Fig.\ref{fig1}.
Taking into account only the $a_{1g}$ band, 
we consider a single band Hubbard model 
$H=\sum_{i,j}\sum_{\sigma}t_{ij}c_{i\sigma}^{\dagger}c_{j\sigma}
+U\sum_{i}n_{i\uparrow}n_{i\downarrow}$, 
 on a 3D triangular lattice,
where the 1st, 2nd and 3rd neighbor in-plane hopping 
integrals $t_1$, $t_2$, $t_3$ and the out-of-plane nearest neighbor 
hopping $t_z$ (Fig.\ref{fig1} upper left) 
are chosen so as to roughly reproduce the $a_{1g}$ portions of the 
band obtained in the first principles calculation.\cite{Singh}
Throughout the paper, we take $t_2=-0.35$, $t_3=-0.07$, and $t_z=-0.15$ 
(unless otherwise noted) in units of $t_1=1$, which corresponds to 
about (or slightly less than) 
0.1eV according to ARPES measurements.\cite{Takeuchi} 
The band (with $U=0$) 
for this choice of parameter values is shown in Fig.\ref{fig1}.
The band filling is $n=$number of electrons/site, 
and it is related to the actual Na content $x$ by $n-1=x$ 
(provided that the $e_g'$ bands are fully occupied). 
We calculate the spin susceptibility 
using the fluctuation exchange (FLEX) approximation\cite{Bickers}
 as $\chi_s(q)=\chi_{\rm irr}(q)/[1-U\chi_{\rm irr}(q)]$, 
where $q=(\Vec{q},i\omega_n)$.
Here, the irreducible susceptibility is 
$\chi_{\rm irr}(q)=-\frac{1}{N}\sum_k G(k+q)G(k)$ 
($N$:number of $k$-point meshes), where $G$ is the renormalized 
Green's function self-consistently 
obtained from the Dyson's equation, in which the 
self energy is calculated using $G$ and $\chi_{\rm irr}$.
We take up to $N=64\times 64\times 64$ $k$-point meshes and 
up to 16384 Matsubara frequencies. 
\begin{figure}[htb]
\begin{center}
\includegraphics[width=8cm,clip]{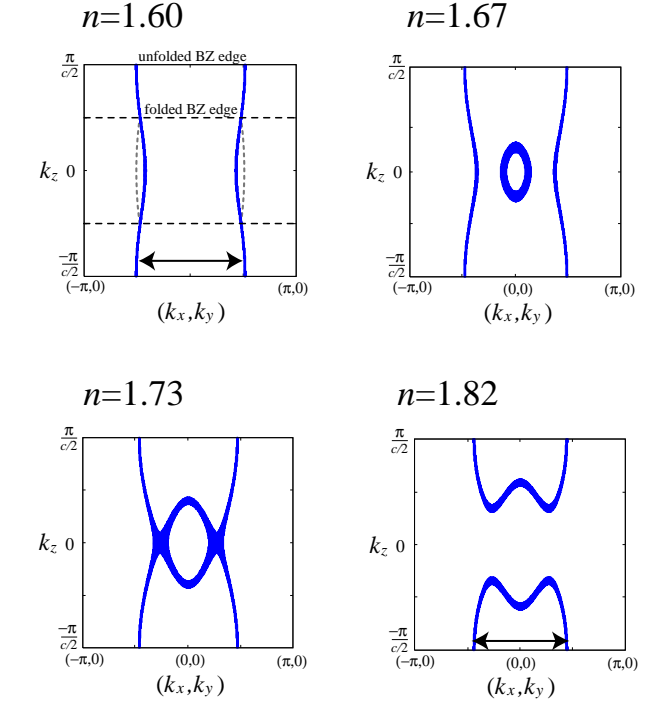}
\caption{(color online) The vertical cross section $(k_y=0)$ 
of the FS for various $n$ with 
$U=0$. Here we use the unfolded BZ scheme, but 
for $n=1.60$ we also show the FS in the usual folded BZ scheme by dashed lines.
The unfolded BZ edge in the $k_z$ direction is denoted 
as ``$\frac{\pi}{c/2}$''  to stress that the unit cell is halved in the 
$c$ direction.
The finite thickness is due to a finite range of energy 
$(E_F\pm 0.03t_1)$
taken in obtaining the FS. The thickness thus represents the density of
states or the effective mass.
The arrows present the diameter of the FS.
\label{fig2}}
\end{center}
\end{figure}
\begin{figure}[htb]
\begin{center}
\includegraphics[width=8cm,clip]{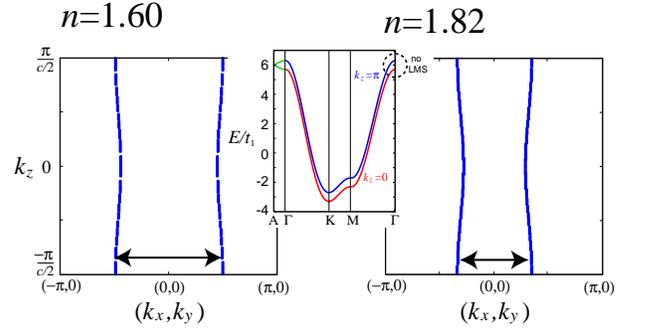}
\caption{(color online) Similar as in Fig.\ref{fig2} except $t_2=t_3=0$. 
The band is shown in the inset.
\label{fig3}}
\end{center}
\end{figure}
\begin{figure}[htb]
\begin{center}
\includegraphics[width=8cm,clip]{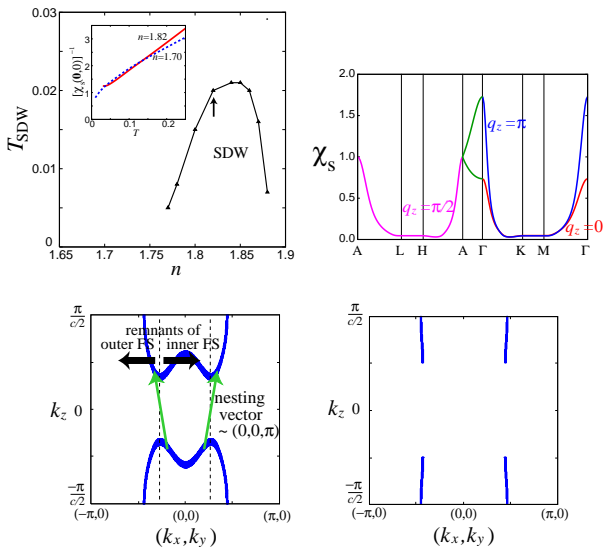}
\caption{(color online) 
Upper left: The band filling dependence of $T_{\rm SDW}$ obtained by FLEX. 
The inset 
shows the $T$ dependence of $\chi_s(\Vec{0},0)^{-1}$ for $n=1.82$ and $n=1.7$.
Upper right: The FLEX $\chi_s(\Vec{q},0)$ for $n=1.82$, $U=6$, 
and $T=0.05$. 
Lower left: the nesting vector of the FS for $n=1.82$ is shown.
Lower right: The FS in the ordered state $(T=0.015)$ obtained by the 
mean field calculation is shown. 
Although the unit cell 
is ``truely'' doubled in this case 
($\pi/2<|k_z|<\pi$ part should 
actually be drawn repeatedly in $0<|k_z|<\pi/2$ region), 
we continue to use the unfolded scheme for 
a clear comparison with the FS in the non-ordered state in the left.
\label{fig4}}
\end{center}
\end{figure}

In order to analyze the magnetically ordered 
state with a wave vector $\Vec{Q}$ at which the FLEX
spin susceptibility is maximized, we consider a 
mean field Hamiltonian
\begin{eqnarray*}
&&H=\sum_{\sigma=\uparrow,\downarrow}\sum_{\Vec{k}}
\sum_{m=0,1}\left[\varepsilon(\Vec{k}+m\Vec{Q})
c^\dagger_{\Vec{k}+m\Vec{Q},\sigma}c_{\Vec{k}+m\Vec{Q},\sigma}\right.\\
&&\left.+U\sum_{l=0,1}\langle n_{m\Vec{Q},-\sigma}\rangle 
c^\dagger_{\Vec{k}+(m+l)\Vec{Q},\sigma}c_{\Vec{k}+m\Vec{Q},\sigma}\right]\\
&&=\sum_{\sigma=\uparrow,\downarrow}\sum_{\Vec{k}}
\sum_{\alpha=1,2}E_{\alpha,\sigma}(\Vec{k})
\gamma^\dagger_{\Vec{k},\sigma,\alpha}\gamma_{\Vec{k},\sigma,\alpha}
\end{eqnarray*}
where $\varepsilon({\Vec{k}})$ is the bare band dispersion,  
$c_{\Vec{k}+m\Vec{Q},\sigma}=\sum_{\alpha=1,2} u_{\alpha,\sigma,m}(\Vec{k})
\gamma_{\Vec{k},\sigma,\alpha}$ is a unitary transformation that diagonalizes 
the Hamiltonian to obtain the two bands $E_{1\sigma}$ and $E_{2\sigma}$, 
and $\langle n_{\Vec{Q},\sigma}\rangle$ is self-consistently determined in the 
usual procedure.
We then calculate the irreducible susceptibility matrix by
\begin{eqnarray*}
&&\left(\chi_0^{\sigma\sigma'}(\Vec{k},\omega)\right)_{ij}
=\frac{1}{N}\sum_{\Vec{p}}\sum_{m,n,\alpha,\alpha'}
u_{\alpha,\sigma,m+i}(\Vec{p}+\Vec{k})\\
&&
\times u_{\alpha,\sigma,m+n+j}(\Vec{p}+\Vec{k})
u_{\alpha',\sigma',m+n}(\Vec{p})u_{\alpha',\sigma',m}(\Vec{p})\\
&&\times 
\frac{f(E_{\alpha',\sigma'}(\Vec{p}))-f(E_{\alpha,\sigma}(\Vec{p}+\Vec{k}))}
{E_{\alpha,\sigma}(\Vec{k}+\Vec{p})-E_{\alpha',\sigma'}(\Vec{p})+\omega
+i\epsilon}, 
\end{eqnarray*}
where $f(E)$ is the Fermi distribution function.
The spin wave dispersion can be determined by the condition that 
the real part of the eigenvalue of the matrix 
$1-U\chi^{\uparrow\downarrow}_0(\Vec{q},\omega)$ equals zero.  
We take up to $N=192\times 192\times 96$ $k$-point meshes here.

We first show in Fig.\ref{fig2} the evolution of the 
3D FS (with $U=0$) as the band filling 
is increased. Here the vertical cross section 
of the FS is given in the 
unfolded  BZ. The FS is a 2D 
cylinder for small band filling, that is, for low Na content. 
As the band filling is increased, an inner FS appears
around the $\Gamma$ point. As in the purely 2D($\sim$ BLH) 
case (see Fig.\ref{fig6} or ref.\onlinecite{Kuroki}), this FS is 
disconnected from the outer 2D cylindrical FS, 
but the inner FS is 3D in the NH case.\cite{comment5}
For larger band fillings, the two FS become connected 
(around $n=1.73$), and then finally for higher band fillings 
it becomes a single 3D FS. 
The appearance of the 3D FS
 despite the small $t_z$ is a consequence of LMS
of the band along the $\Gamma$-A line.
In fact, if we take a band 
that does not have LMS, $i.e.,$ for $t_2=t_3=0$ but with the 
same $t_z(=-0.15)$, the 
FS remains to be 2D 
(unless the band filling becomes very close to 2) as shown in Fig.\ref{fig3}.
The 3D evolution of the FS is consistent  with  the 
ARPES observations, where the in-plane $k_F$ barely 
decreases with the increase of the Na content 
for $x>0.6$.\cite{Yang2,Hasan2} In fact, the decrease of the 
FS diameter in Fig.\ref{fig2} is very small compared to that  
in Fig.\ref{fig3}. 


In the inset of the upper left of Fig.\ref{fig4} ,
FLEX results of $\chi^{-1}_s(\Vec{q}=0,i\omega_n=0)$ 
are plotted as functions of $T$ 
for $U=6$ and $n=1.82$ or $n=1.7$. 
The Curie-Weiss behavior  at 
high temperatures with a negative Curie-Weiss temperature is 
consistent with the experiments.
\cite{Motohashi,Sugiyama,Galivano,Keimer1,Foo,Wang,Imai,Alloul}
In the upper left of Fig.\ref{fig4}, $T_{\rm SDW}$, defined 
as the temperature where the maximum value of $U\chi_{\rm irr}$ reaches 
$\simeq 1$, is plotted as a function of the band filling. 
$T_{\rm SDW}$ exists only in the 
regime $1.75<n<1.9$, which is consistent with the fact that the 
magnetic ordering is not observed experimentally for $x<0.75$.\cite{Foo,Imai} 
The maximum $T_{\rm SDW}$ corresponds to about 20K, in agreement
with the experiments.\cite{Motohashi,Sugiyama,Keimer1}
Around $n\sim 1.8$, the ordering occurs with $\Vec{Q}=(0,0,\pi)$, as shown in 
the upper right of Fig.\ref{fig4} for $n=1.82$, which corresponds to 
in-plane ferromagnetic, 
out-of-plane antiferromagnetic spin structure,
as found experimentally.\cite{Keimer2,Helme} 
The origin of this spin correlation can be found in the 
shape of the 3D FS. 
Namely, the FS at $n\sim 1.8$ 
is partially nested with a nesting vector close to 
$(0,0,\pi)$, as can be seen in Fig.\ref{fig4} (lower left panel). 
In fact, this nesting occurs 
between portions of the FS that 
can be considered as remnants of the 
inner and outer FS, which are disconnected 
for lower band fillings.\cite{comment2} 
On the other hand, 
as the band filling comes close to $n=1.9$ ($x=0.9$), 
the FS becomes too small for $(0,0,\pi)$ nesting, and the 
ordering wave number becomes incommensurate in the $c$ direction, 
which is also consistent with the finding in ref.\onlinecite{Sugiyama}. 
\begin{figure}[b]
\begin{center}
\includegraphics[width=7cm,clip]{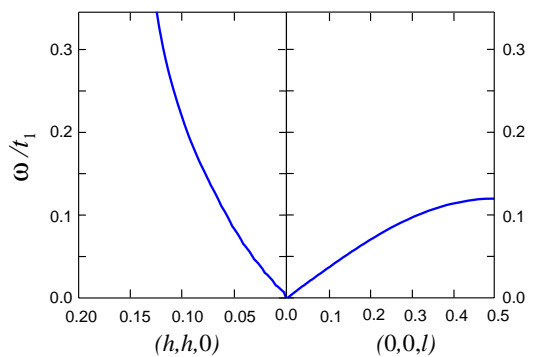}
\caption{(color online) 
The calculated in-plane (left) and out-of-plane (right)
spin wave dispersion for 
$n=1.82$, $U=4$, 
and $T=0.015$.
Here the wave vectors are presented in units of the reciprocal 
lattice primitive vector 
in the usual {\it folded} BZ for clear comparison with 
Fig.4 of ref.\onlinecite{Keimer2}. 
\label{fig5}}
\end{center}
\end{figure}
\begin{figure}[htb]
\begin{center}
\includegraphics[width=8cm,clip]{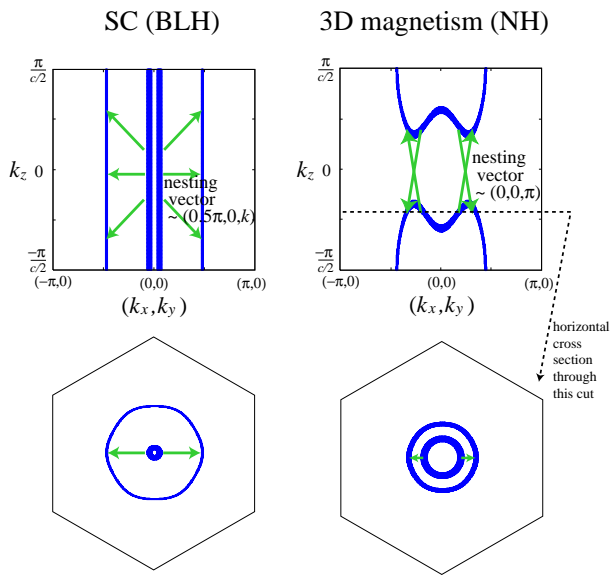}
\caption{(color online) The FS nesting 
relevant for SC and the 3D magnetism.
\label{fig6}}
\end{center}
\end{figure}

We now move on to the mean field results 
in the $\Vec{Q}=(0,0,\pi)$ SDW ordered state.
We focus on $n=1.82$ and take $U$ smaller than $U=6$ adopted in FLEX 
since the mean field 
approach tends to overestimate the tendency toward ordering.
In fact, taking $U=4$ gives a self-consistently determined 
$\langle n_{\Vec{Q}\uparrow}\rangle$ value of 0.075 at low $T$, 
corresponding to a magnetic moment of $0.15\mu_B$ per site (Co atom), 
in rough agreement with a $\mu$SR estimation $0.18\mu_B$.\cite{Sugiyama}
In Fig.\ref{fig4} (lower right panel), 
we show the calculation result of the FS in the ordered state.  
The portions of the FS around $k_z=\pm\pi/2$ 
disappear, but other portions remain to result in a 
FS with a strong 2D character. 
This is consistent with the fact that 
system remains metallic in the magnetically ordered state.\cite{Motohashi}
Note also that the disappearing portions of the 
FS are thick, i.e., they have heavy mass,  
which is consistent with what has been suggested in ref.\onlinecite{Motohashi}
from the enhancement of the mobility in the magnetically ordered state.

In Fig.\ref{fig5}, we show the calculated  spin wave 
dispersion. The overall feature as well as the in-plane to out-of-plane 
ratio $\omega(0.1,0.1,0)/\omega(0,0,0.5)\sim 2.1$ strikingly resembles 
the neutron scattering result of ref.\onlinecite{Keimer2} for $x=0.82$.
The energy scale ($\omega(0,0,0.5)\simeq 0.1t_1$ 
corresponds to $\sim 10$ meV) 
is also close to the experimental result.
Here again, 
our results show that the 3D magnetism is well understood within the 
present itinerant spin picture.\cite{Sushko}


Finally, let us discuss the relation 
between this 3D magnetism 
and SC. 
As we have seen, the SF in the 
nonhydrated systems becomes weak for small band fillings.
This is because the nesting between the 2D outer and the 
3D inner FS is not good.
However, the nesting between the inner and the outer FS\cite{comment3} can be 
restrengthened if the two dimensionality 
is increased by hydration, resulting in enhanced SF.
As we have mentioned in the beginning,
our proposal is that 
this SF is responsible for SC.\cite{Kuroki}
Thus, in our view, the 3D magnetism and 
the incommensurate SF that gives rise to SC 
share the same origin as shown in Fig.\ref{fig6}: 
the nesting between the inner and the outer 
(connected or disconnected) FS that originate from 
the local minimum of the $a_{1g}$ band.

Numerical calculation were performed
at the facilities of the Supercomputer Center,
ISSP, University of Tokyo.
This study has been supported by 
Grants-in-Aid for Scientific Research from the Ministry of Education, 
Culture, Sports, Science and Technology of Japan, and from the Japan 
Society for the Promotion of Science.

%


\begin{thebibliography}{99}
\bibitem[*]{SO} Present affiliation: NHK Spring Co.
\bibitem[**]{TN} Present affiliation: Nikon Co.
\bibitem{Takada} K. Takada {\it et al.}, 
Nature {\bf 422}, 53 (2003).
\bibitem{Hasan} M.Z. Hasan {\it et al.}, 
Phys. Rev. Lett. {\bf 92}, 246402 (2004).
\bibitem{Yang2} H.-B. Yang {\it et al.}, Phys. Rev. Lett. 
{\bf 95}, 146401 (2005).
\bibitem{Takeuchi} T. Takeuchi {\it et al.}, 
Proc. 24th Int. Conf. Thermoelectrics. (2005) p.435.
\bibitem{Shimojima} T. Shimojima {\it et al.}, cond-mat/0606424.
\bibitem{Singh} D.J.Singh, Phys. Rev. B {\bf 61}, 13397 (2000).
\bibitem{Ishida3} K. Ishida {\it et al.}, 
J. Phys. Soc. Jpn. {\bf 72}, 3041 (2003).
\bibitem{Fujimoto} T. Fujimoto {\it et al.}, 
Phys. Rev. Lett. {\bf 92}, 047004 (2004).
\bibitem{Imai} F.L. Ning {\it et al.}, 
Phys. Rev. Lett. {\bf 93}, 237201 (2004); F.L. Ning and 
T. Imai, Phys. Rev. Lett. {\bf 94}, 227004 (2005).
\bibitem{Ihara} Y. Ihara {\it et al.},
J. Phys. Soc. Jpn. {\bf 74}, 867 (2005).
\bibitem{Michioka} C. Michioka {\it et al.}, J. Phys. Soc. Jpn. {\bf 75}, 
063701 (2006).
\bibitem{Ihara2} Y. Ihara {\it et al.}, 
J. Phys. Soc. Jpn. {\bf 74}, 2177 (2005).
\bibitem{Alloul} I.R. Mukhamedshin {\it et al.}, Phys. Rev. Lett. {\bf 94}, 
247602 (2005).
\bibitem{Kobayashi} Y. Kobayashi {\it et al.}, 
J. Phys. Soc. Jpn. {\bf 74}, 1800 (2005).
\bibitem{Zheng} G.-q. Zheng {\it et al.}, J. Cond. Matt. {\bf 18}, L63 (2006).
\bibitem{Zheng2} G.-q. Zheng {\it et al.}, Phys. Rev. B {\bf 73}, 180503 
(2006).
\bibitem{Yokoi} M. Yokoi {\it et al.},
J. Phys. Soc. Jpn. {\bf 73}, 1297 (2004).
\bibitem{Oeschler} N. Oeschler {\it et al.}, cond-mat/0503690.
\bibitem{Kuroki} K. Kuroki {\it et al.}, Phys. Rev. B {\bf 75}, 051013 (2006).
\bibitem{MO} M. Mochizuki and M. Ogata, cond-mat/0609443, to be published in 
J. Phys. Soc. Jpn.
\bibitem{Motohashi} T. Motohashi {\it et al.}, Phys. Rev. B {\bf 67}, 064406
(2003).
\bibitem{Sugiyama} J. Sugiyama {\it et al.}, Phys. Rev. B {\bf 67}, 214420 
(2003); {\it ibid.} {\bf 69} 214423 (2004).
\bibitem{Foo} M.L. Foo {\it et al.}, Phys. Rev. Lett. {\bf 92}, 247001 (2004).
\bibitem{Keimer1} 
S.P. Bayrakci {\it et al.}, Phys. Rev. B {\bf 69}, 100410 (2004).
\bibitem{Keimer2} 
S.P. Bayrakci {\it et al.}, Phys. Rev. Lett. {\bf 94}, 157205 (2005).
\bibitem{Helme} A.T.Boothroyd {\it et al}, Phys. Rev. Lett. {\bf 92}, 197201 
(2004); 
L.M. Helme {\it et al.}, Phys. Rev. Lett. {\bf 94}, 157206 (2005).
\bibitem{Johannes} M.D. Johannes {\it et al.}, Phys. Rev. B {\bf 71}, 
214410 (2005).
\bibitem{Galivano} J.L. Galivano {\it et al.}, Phys. Rev. B {\bf 69}, 100404 
(2004).
\bibitem{Wang} Y. Wang {\it et al.}, Nature {\bf 423}, 425 (2003).
\bibitem{Khaliullin} M. Daghofer {\it et al.}, Phys. Rev. Lett. {\bf 96}, 
216404 (2006).
\bibitem{Korshunov} Quite recently, M.M. Korshunov {\it et al.} in 
cond-mat/0608327 discuss the relation between LMS and ferromagnetic 
SF, but neglect the three dimensionality of the system.
\bibitem{Bickers} N.E. Bickers, D.J. Scalapino, and S.R. White, 
Phys. Rev. Lett. {\bf 62}, 961 (1989).
\bibitem{comment5} The term ``inner'' and ``outer'' FS in the present 
study should not be mixed up with the commonly known 
``bilayer-coupling-originated inner and outer''   
$a_{1g}$ FS, which in the unfolded BZ scheme correspond to FS 
at $k_z=0$ and $k_z=\frac{\pi}{c/2}$ (see Fig.\ref{fig2}, $n=1.60$).
\bibitem{Hasan2} D. Qian {\it et al.}, Phys. Rev. Lett. {\bf 96}, 216405 
(2006).
\bibitem{comment2} We have investigated other sets of 
hopping values 
and found that the $(0,0,\pi)$ spin correlation (with a possible 
slight incommensurability as expected from the nesting vector) is 
robust when the band filling is 
around $n\sim 1.8$  as far as LMS of the band exists.
\bibitem{Sushko} 
In cond-mat/0509308, Y.V. Sushko {\it et al.} find that   
the magnetic ordering temperature 
increases with applying pressure, unusual for an SDW state.
In fact, applying pressure should increase $|t_z|/t_1$, which 
according to our FLEX calculation  
{\it favors} the SDW (as far as the increase of $|t_z|$ is within $\sim 10\%$) 
because the origin of the nesting here is the 3D FS itself.
Applying pressure may also increase 
$|t_2|/t_1$ and $|t_3|/t_1$,
which makes LMS of the band deeper, and thus also favors the SDW. 
\bibitem{comment3} The inner FS can appear despite the low Na content of 
$x\sim 0.35$ because of the presence of H$_3$O$^+$ ions suggested in e.g.,
H. Sakurai {\it et al.}, Phys. Rev. B {\bf 74}, 092502 (2006).


\end{thebibliography}
\end{document}